\documentclass[]{preprint}
\pdfsuppressptexinfo=-1 \pdfinfoomitdate=1 \pdftrailerid{}
\usepackage[utf8]{inputenc}
\usepackage{inconsolata}
\hypersetup{pdftitle={Using the Tool That Replaces You: Generative AI and the Adoption Bind among Bangladeshi Platform Freelancers},pdfauthor={Naim Bin Hasan},pdfcreator={},pdfproducer={}}

\title{Using the Tool That Replaces You: Generative AI and the Adoption Bind among Bangladeshi Platform Freelancers}
\author{Naim Bin Hasan}
\affiliation{School of Information Studies, Syracuse University, Syracuse, New York, USA}
\abstract{Platform studies show that generative AI has reduced demand for freelance writing and image work, and experiments show that it makes individual workers faster. This letter asks how these two findings look to workers who are both users of the technology and exposed to it. I analyze 30 interviews with Bangladeshi online freelancers, conducted in Bangla between May and August 2025. Twenty-one participants used generative AI in their own work, sixteen said that it was replacing freelance work, and thirteen said both. Eleven of the sixteen placed replacement on others: on beginners and workers who do not update their skills, on ``basic'' tasks, or on other sectors. Participants also described how prices fell as tasks became faster, and how clients cited AI to offer less. I propose the term adoption bind for this pattern: each worker adopts generative AI to stay on the safe side of the line, while the same tools, in the hands of competitors and clients, lower what the market pays for the work. The bind suggests that productivity gains measured for individuals can become price losses for workers as a group, and that these losses fall first on entry-level work.}
\checkdata[\faTags\;Keywords]{generative AI; platform labor; online freelancing; Global South; Bangladesh; pricing}
\correspondence{\email{naimbhdu@gmail.com}}
\checkdata[ORCID]{\href{https://orcid.org/0009-0002-4930-7241}{\texttt{0009-0002-4930-7241}}}
\date{November 2026}

\begin{document}
\maketitle
\section{Introduction}

Two bodies of evidence now describe how generative AI affects freelance work. The first uses platform data. On a leading freelancing platform, job posts for automation-prone writing and coding work fell by 21\% relative to manual-intensive work within eight months of ChatGPT's release, and posts for image creation fell by 17\% after image generators arrived \citep{demirci2025}. Freelancers in the most affected occupations lost both jobs and earnings \citep{hui2024}, and demand shifted between skills, producing clear winners and losers \citep{teutloff2025}. The second body uses experiments and firm data. ChatGPT cut the time that college-educated professionals spent on writing tasks by 40\% and raised the quality of their output by 18\% \citep{noy2023}. An AI assistant raised the productivity of customer-support agents by 15\% on average, with the largest gains among less experienced workers \citep{brynjolfsson2025}.

These studies place workers on one side of the technology: as people whose demand falls or as people whose output rises. Many platform freelancers stand on both sides at once, because they use the same tools that clients can use instead of hiring them. Studies of how workers themselves see this change come mostly from the Global North. In workshops with 54 knowledge workers in three US cities, participants largely saw generative AI as a tool for menial work under human review and did not expect it to disrupt their industries \citep{woodruff2024}. Workers in the Global South who sell that kind of work on global platforms face a different position. They compete for clients in richer countries in a market with far more workers than jobs \citep{graham2017}, and the freedom of platform work comes with high precarity \citep{anwar2021}.

This letter asks one question: How do Bangladeshi freelancers who use generative AI in their own work understand its effect on the market for that work? I answer it with interviews with 30 freelancers, and I propose one concept, the adoption bind, for the pattern they describe.

\section{Study}

I interviewed 30 Bangladeshi online freelancers between May and August 2025 as part of a larger study of platform work. Participants were recruited through personal networks, referrals and Bangladeshi freelancer groups on Facebook. Each had completed at least three paid projects, had earned more than US\$300 and held an active profile on a platform such as Upwork or Fiverr. Twenty-one were men and nine were women, and 25 were students. By main sector, 11 worked in design and media, nine in software development and nine in marketing and writing. One did small website and posting tasks. I conducted all interviews myself. They lasted 45 to 70 minutes, took place on Zoom or Google Meet in Bangla and were audio-recorded with verbal consent. My own three years as a platform freelancer helped rapport, but it also meant that I checked my readings against what participants actually said. The study was approved by the Institutional Review Board of Florida Atlantic University (protocol IRB2502046). All names are pseudonyms.

Each interview asked whether artificial intelligence had affected the participant's work and how it might change freelancing. Participants used the English word ``AI'' within Bangla speech. The tools they named, such as ChatGPT, Gemini, Canva and AI photo editors, show that they meant generative AI. I coded all 96 answers that discussed AI with a codebook of 12 codes, such as \emph{uses AI tools}, \emph{AI replaces work}, \emph{price pressure} and \emph{client expectations}. A language model made a first pass. I then checked every answer against the transcript and changed the codes of six answers. The first pass and my final codes agreed closely (Cohen's $\kappa = 0.98$). This figure shows how much I changed the model's work, not agreement between independent human coders, and every final code is my own judgment. I count a participant under a code if any of their answers carries it. For this letter, I then reread the 31 answers coded \emph{AI replaces work} and noted whether each participant placed replacement on a group, task or sector apart from their own work. These counts describe this sample, not the share of Bangladeshi freelancers who hold a view. I translated all quotations from the Bangla transcripts.

\section{Findings}

\subsection{Using the tool and expecting it to replace}

Twenty-one of the 30 participants used generative AI in their freelance work, and sixteen said it was replacing tasks, jobs or demand for human freelancers. Thirteen were in both groups (Table~\ref{tab:codes}). Use was most common in design and media, where 10 of the 11 participants used the tools.

\begin{table}[t]
\caption{Participants whose answers carried each code, by main sector. A participant is counted once per code.}
\label{tab:codes}
\begin{tabular}{lcccc}
\toprule
Code & All & \shortstack{Design and\\media} & Development & \shortstack{Marketing\\and writing} \\
 & (30) & (11) & (9) & (9) \\
\midrule
Uses AI tools & 21 & 10 & 5 & 6 \\
AI replaces work & 16 & 7 & 5 & 4 \\
\quad Also uses AI tools & 13 & 6 & 4 & 3 \\
\quad Places replacement on others & 11 & 4 & 4 & 3 \\
Must change skills & 10 & 3 & 4 & 3 \\
Client expectations & 4 & 2 & 1 & 1 \\
Price pressure & 3 & 1 & 0 & 2 \\
\bottomrule
\end{tabular}
\par\smallskip\footnotesize Indented rows are subsets of \emph{AI replaces work}. One participant who did small website and posting tasks carried none of these codes.
\end{table}

The users described real gains. Mehedi Hasan Limon, a 26-year-old business graduate who works in digital marketing and design and runs a digital marketing agency, used Gemini and ChatGPT Plus every day: ``Through this I have been able to increase my productivity even more. The work now takes much less time than it used to.'' Meem Jahan, a 24-year-old student who builds websites, said: ``I do not currently write 40 percent of my code myself. I use ChatGPT Plus. It gives me some boilerplate, and then I edit it and add colors and pictures and so on.'' Mostafizur Rahman, a 25-year-old undergraduate and AI user who works in graphic design and video editing, described what the same speed meant for the market: ``In graphic design, a logo that used to take us 3 to 5 days to create, now AI can create it in at most 15 minutes. And due to that, many are losing their jobs.''

\subsection{Placing replacement on others}

Eleven of the sixteen who described replacement placed it on others, below or beside themselves. Five drew the line by skill. Mehedi Hasan Limon asked and answered the question in one breath: ``So whom will AI replace? Those who cannot do anything. Those who are fairly good can use AI as a tool and increase their productivity.'' Imran Mahmud, a 21-year-old diploma engineering student who develops apps, separated experts from beginners: ``Those who are experts, who keep themselves updated day by day, for them, AI is not bad. [\dots] But the ones who are beginners or new, for them, it has become a bit difficult.'' Faisal Chowdhury, a 23-year-old web developer who works in a software company, saw the same line inside firms: ``Now there are no junior positions; they have all been replaced by AI.'' In his account, one senior graphic designer can now do the work of four people. Three placed replacement on basic tasks, such as banners, short essays and simple logos, and kept the harder tasks for humans. Three placed it on other sectors. Enamul Haque, a 25-year-old student and web developer, said that AI had a ``very bad effect on graphic designing or content writing,'' but that web work was still in high demand.

The other five did not exempt themselves. They spoke of what ``we'' used to do. Emon Shikder, a 21-year-old student who does digital marketing and voice-over work, said: ``What takes us a lot of time, it can do very quickly. So in that case, AI feels like a big problem.'' For the eleven, the line works as a kind of insurance, because using AI and updating one's skills is how a worker stays on the safe side. Ten participants said that AI forced them to change or update their skills.

\subsection{When the safe side sets the price}

Participants also described a market that reprices the task itself. Sajjad Hossain, a 24-year-old student who had moved from graphic design to academic writing, explained the move: ``The basic tasks that are now possible with AI in two minutes, we used to get paid a good amount for those. So, lots of the basic work is gone now. This is the big reason I switched my work sector.'' He described an unwritten rule within his community of freelancers that nobody would design a logo for less than five dollars. That rule had collapsed: ``now people are designing logos for one, two, three dollars.'' He blamed this first on new, inexperienced freelancers who underbid, and then on template platforms such as Canva and Pixlr and on AI. He summed up the result in one sentence: ``The work has become easier, and the easier the work became, the more the pricing came down.'' Afsana Islam, a 23-year-old business student who does graphic and web design, put it more starkly. Small tasks such as images and content ideas, she said, are ``now normally available for free.''

Clients used the same reasoning in negotiation. Israt Jahan, a 30-year-old MBA graduate who does search engine optimization and social media marketing while caring for her sick mother, described it this way: ``When I offer a design or a project, the client may offer a lesser price for it. When I ask why, they say, `I can do this with AI, too.'\,'' Israt's client did not need AI to do the job. The claim that AI could do it was enough to lower the offer. Jakir Khan, a 25-year-old student who runs an agency for video, marketing and web work, said that graphic design ``used to have a lot of value, a lot of price, but now the value of graphic design is much lower because AI has replaced it.''

Clients also changed what they hired for. Lamia Siddiqua, a 26-year-old student and graphic designer, said that many clients now ``generate an idea for themselves from AI'' and send it as the brief, which she then has to redo ``in our own way.'' Faisal Chowdhury said that clients often build ``60 to 70\%'' of a software project with AI tools and then cannot move forward. Handling the rest, he said, takes a fairly senior developer. In both cases the client keeps the cheap first draft and pays a human only for the smaller, harder remainder. Three participants described price pressure directly and four described changed client expectations, so these accounts show a mechanism, not its frequency.

\section{Discussion: The Adoption Bind}

I propose the term \emph{adoption bind} for this pattern. Each worker adopts generative AI to stay on the safe side of the line between those who use AI and those it replaces. But the same tools, in the hands of competitors and clients, make the work look easier and lower what the market pays for it. The interviews show the two halves of the bind side by side, not a measured causal link between them. No participant said that their own use of AI had lowered their prices.

The concept connects the two bodies of evidence in the introduction. The productivity gains that experiments measure for individuals \citep{noy2023,brynjolfsson2025} were real for the participants who used AI. When competitors and clients hold the same tools, however, the time a task takes falls for everyone, and, in these accounts, the price follows. \citet{acemoglu2019} call the general version of this the displacement effect of automation, which moves tasks from labor to capital and can reduce labor demand even as productivity rises. The interviews add the small step through which displacement reaches a single freelancer: a client's claim, in one negotiation, that AI could do the task. They also differ from the US knowledge workers who expected generative AI to take only menial work \citep{woodruff2024}. Participants here agreed about menial work, but for them that work was already part of what they sold.

The bind has three implications. First, it falls hardest on entry-level work. Participants placed replacement on beginners and basic tasks, and platform data show the largest falls in writing and image creation \citep{demirci2025,hui2024}. If these tasks disappear, the line that experienced workers draw also closes the route by which beginners learn enough to cross it. This matters for workers like those in this study: 25 of the 30 were students, and 18 supported their families fully or in part. Second, staying on the safe side requires unpaid learning. \citet{blaising2021} describe the financial, emotional, relational and reputational overhead of keeping an online freelance career. Generative AI adds a technical layer to that overhead, and freelancers pay for it alone. Third, measures of exposure built from the tasks that AI can perform \citep{eloundou2024} capture only part of the effect. Israt's case shows that a task can lose value through a client's belief about what AI can do, whether or not AI does the task well.

Platform data could test the bind by tracking two measures: the price offered for the same task category over time, which would show whether prices follow faster task times, and the share of jobs won by new accounts, which would show whether entry routes are closing. Interviews with clients could test whether claims about AI lower their offers.

This study has clear limits. It rests on 30 interviews, mostly with young students recruited through networks and Facebook groups. The data are self-reports from mid-2025, a period of fast change in the tools. Participants' accounts of clients are second-hand. The codes, including the reading of who placed replacement on others, are my own, with no second human coder. The adoption bind is therefore a proposition to test, not a measured effect.

\section{Conclusion}

Most Bangladeshi freelancers in this study used generative AI, and most of those users also expected it to shrink their market. Most of them reconciled the two by placing replacement on others. Their accounts suggest that this answer may not hold for long, because when workers and clients use the same tools, the faster time becomes the basis for the price. Research on generative AI and work should therefore study what clients pay, not only how fast workers produce.

\section*{Acknowledgments}
I thank the 30 freelancers who shared their time and experience. This research was supported by an American Institute of Bangladesh Studies (AIBS) Fellowship and a Florida Atlantic University Graduate Thesis and Dissertation Scholarship.

\section*{Data Availability}
The interview recordings and transcripts are not publicly available, because participants were told at the start of each interview that the recordings and any identifying information would not be shared. The codebook and participant-level code counts are available from the author on request.

\section*{GenAI Usage Disclosure}
A large language model translated the Bangla transcripts into English for analysis and made a first pass of coding, which the author checked and corrected answer by answer, as described in Section~2. Some quotations were translated from the Bangla originals with AI assistance, and the author checked each one against the original. The author checked every claim, count, quotation and reference against the data and the sources and takes full responsibility for the content.

\bibliographystyle{plainnat}
\bibliography{references}
\end{document}